\documentclass[twocolumn,showpacs,preprintnumbers,amsmath,amssymb,amsbsy,showkeys]{revtex4}

\usepackage{graphicx}% Include figure files
\usepackage{dcolumn}% Align table columns on decimal point
\usepackage{bm}% bold math
\usepackage{bezier}

\begin{document}

\title{Pulsed high harmonic generation of light due to pumped Bloch 
oscillations in 
noninteracting metals}
\author{J. K. Freericks$^1$, A. Y. Liu$^1$, A. F. Kemper$^{2,3}$, and T. P. Devereaux$^{2,3}$}
\affiliation{$^1$Department of Physics, Georgetown University, 37th and O Sts. NW, Washington, DC 20057, USA\\
$^2$Stanford Institute for Materials and Energy Science,
SLAC National Accelerator Laboratory, Menlo Park, CA 94025, USA\\
$^3$Geballe Laboratory for Advanced Materials,
Stanford University, Stanford, CA 94305, USA
}%

\date{\today}% It is always \today, today,

\begin{abstract}

We derive a simple theory for high-order harmonic generation
due to pumping a noninteracting metal with a  large amplitude 
oscillating electric field.  The model assumes that the radiated 
light field arises from the acceleration of electrons due to the 
time-varying current generated by the pump, and also assumes that the 
system has a constant density of photoexcited carriers, hence it 
ignores the dipole excitation between bands (which would create 
carriers in semiconductors).  We examine the circumstances  under which 
odd harmonic frequencies would be expected to dominate the spectrum
of radiated light, and we also apply the model to real materials like ZnO, 
for which high-order harmonic generation has already been demonstrated 
in experiments. 

\end{abstract}
\maketitle

\section{Introduction}
\label{sec:Introduction}

The theory of Bloch oscillations has a long history, dating back to the
original work by Bloch~\cite{bloch} and Zener~\cite{zener}.  It remains one
of the most difficult phenomena to observe in real materials due to the fast
timescale for the oscillations, and it has been seen predominantly in
semiconductor superlattices~\cite{superlattice} and in ultracold atomic
systems when placed on optical lattices~\cite{optical_lattice}. We argue here
that similar phenomena can be observed by examining the radiated light from 
solids that are pumped by high intensity laser pulses with femtosecond 
pulse widths~\cite{kemper, golde}.

We describe such laser pulses as spatially-uniform time-varying electric fields,
using a semiclassical approach.  In this sense, we ignore all magnetic field
effects, and essentially violate Maxwell's equations for the electromagnetic
fields.  This approach will be accurate if the magnetic field effects on the 
solid are small, which should be true for rapidly oscillating fields, because 
the electron spin will not be able to respond to the magnetic field 
on such short time scales.  The large electric field will cause the electrons
in the material to accelerate, giving rise to a time-dependent current.
Assuming the current on the surface is proportional to the bulk current, 
we will see electromagnetic radiation coming off the material with a signal
proportional to the time derivative of the current.  We will use this to directly
determine the spectra of the radiated light.

Experimental investigations of high harmonic generation in 
bulk crystals are being pursued, including recent work on
irradiating ZnO crystals, where many tens of harmonics
of the mid-infrared driving pulse were measured~\cite{zno_exp}. 
For most crystallographic orientations, only the odd harmonics of the 
fundamental frequency appeared.
We will see below that the occurence of only odd harmonics
is to be expected for systems that satisfy time-reversal symmetry.

In the next section, we discuss the general formalism, and apply it to simple
model systems, which can be thought of as tight-binding models for the 
bandstructure. In Section III, we use numerics to examine the high harmonic generation in simple models, and in Section IV, we use numerics to examine the high harmonic generation
in ZnO and compare the numerical results with those of experiment.
We conclude in the final section.

\section{Formalism}

We begin with a system described by a bandstructure, which we denote by
$\epsilon({\bf k})$. The bandstructure is a periodic function of the 
Brillouin zone.  If we use a tight-binding description for the bands, then
the bandstructure can generically be written as
\begin{equation}
\epsilon({\bf k})=-\sum_{\boldsymbol \delta} t_{\boldsymbol \delta} e^{i{\bf k}\cdot{\boldsymbol \delta}},
\label{eq: band_structure}
\end{equation}
for $s$-orbitals, where $\boldsymbol \delta$ is a translation vector to a neighboring site
(the sum need not stop at just nearest neighbors), and $-t_{\boldsymbol \delta}$ is the corresponding
hopping integral.  The band index, if there are multiple bands, has been suppressed. For models with higher angular momentum
orbitals, the tight-binding form for the bandstructure
is more complicated and requires diagonalization of a matrix whose size
is equal to the number of different types of orbitals per
unit cell (times two in the presence of magnetic or spin-orbit 
interactions). 

We work in the Hamiltonian gauge, where the electric field ${\bf E}(t)$
is described by a spatially uniform vector potential ${\bf A}(t)$, via
\begin{equation}
{\bf E}(t)=-\frac{1}{c}\frac{\partial {\bf A}(t)}{\partial t},
\label{eq: efield}
\end{equation}
with  $c$ the speed of light.
The electric field is then introduced into the bandstructure with the so-called
Peierls substitution which takes $\epsilon({\bf k})\rightarrow \epsilon({\bf k}-e{\bf A}(t)/\hbar c)$, with $e<0$ the electron charge.  This substitution properly describes the way in
which electrons are accelerated in each band, but does not take into account the 
dipole coupling between different types of bands (like between $s$ and $p$
bands). Such a direct Zener tunneling term can be included, but will
complicate the analysis, and we ignore it in the results we
describe here, for simplicity.  We expect that the results one finds if
we include such terms will be similar, because the main effect of the
laser pulse is to accelerate the electrons within a given band in metallic
systems, and excitations between bands become more important for semiconductors
or insulators.  But even in that case, if we assume most of the photoexcitation
occurs when the field amplitude is the highest, then the results of this work
will be identical except for a factor of 2, which does not affect any of the 
results we present below.

The Hamiltonian of the system then becomes
\begin{equation}
\mathcal{H}(t)=\sum_{{\bf  k}\sigma}\epsilon\left ( {\bf k}-\frac{e{\bf A}(t)}{\hbar c}\right )
c_{{\bf k}\sigma}^{\dagger}c_{{\bf k}\sigma}^{}
\label{eq: ham1}
\end{equation}
where $c_{{\bf k}\sigma}^{\dagger}$ ($c_{{\bf k}\sigma}^{}$) is the electron creation (annihilation) operator for an electron with momentum ${\bf k}$ and spin $\sigma$ (here we are assuming, for simplicity, a single band model and no spin-orbit coupling, so $\sigma$ is a good quantum number).  This is a time-dependent Hamiltonian, but it commutes with itself at different times, which makes its time evolution particularly easy to solve.

The system starts off in equilibrium at early times $t\rightarrow -\infty$ and the electric field pulse is centered around $t=0$. Hence, the initial distribution of the electrons is given by the Fermi-Dirac distribution
\begin{equation}
f(\epsilon-\mu)=\frac{1}{1+\exp[\beta(\epsilon-\mu)]}
\label{fermi_dirac}
\end{equation}
in the distant past.
Here, $\mu$ is the chemical potential, and $\beta$ is the inverse temperature. We work with an isolated system, so the time dynamics can be solved directly in the Heisenberg picture. The current satisfies
\begin{equation}
{\bf J}(t)=\frac{e}{\hbar}\sum_{{\bf k}\sigma}{\boldsymbol \nabla}_{\bf k}\epsilon\left ( {\bf k}-\frac{e{\bf A}(t)}{\hbar c}\right ) f(\epsilon_{\bf k}-\mu),
\label{eq: current}
\end{equation}
which simply sums the field-dependent velocity ($\nabla_{\bf k}\epsilon /\hbar$), weighted by the initial distribution of the electrons and multiplied by the electric charge. This formula also follows from a Kadanoff-Baym-Keldysh formalism~\cite{freericks_nonint} where one can exactly solve for the Green's functions, evaluate them at equal time and sum the electron occupancy at ${\bf k}$ weighted by the field-dependent velocity.

If we assume that the radiated light arises from the acceleration of the electrons, then the light spectrum is proportional to
\begin{equation}
\left | \int dt e^{i\omega t} \frac{d}{dt}{\bf J}(t)\right |^2 =\left | \omega\int dt e^{i\omega t}{\bf J}(t)\right |^2 .
\end{equation}
If we have a pulse with a shape given by 
\begin{equation}
{\bf A}(t)=-{\bf A}_0\cos(\omega_0 t+\phi)\frac{\exp^{(-t^2/t_0^2)}}{t_0\sqrt{\pi}},
\end{equation}
then we are guaranteed that the field has no dc component (required of optical pulses) and if $t_0$ is somewhat larger than the period of the oscillations $2\pi/\omega_0$, then the pumped electric field has a reasonably well defined fundamental frequency ($\phi$ is an adjustable phase shift of the oscillations relative to the gaussian envelope). While we choose a gaussian form for the 
envelope function,  any normalized slowly varying function which vanishes rapidly enough as $|t|\rightarrow\infty$ can be used for the envelope, which we denote below by $g(t)$. By calculating the spectrum of the radiated light from the Fourier transform of the current as a function of time, we can examine the high harmonic generation in the system.

Before jumping into numerics, we first discuss some general principles and examine symmetry considerations. If the lattice is inversion symmetric, then the parity operation immediately tells us that the energy is an even function of the wavevector
\begin{equation}
\epsilon({\bf k})=\epsilon({-\bf k}).
\label{eq: energy_parity}
\end{equation}
If the lattice lacks inversion symmetry, this result still holds if there is 
no spin-orbit coupling and no magnetic field due to magnetic order.  
This is because the Hamiltonian is time-reversal symmetric, so there is 
a Kramers doublet, 
\begin{equation}
\epsilon_\uparrow({\bf k})=\epsilon_\downarrow(-{\bf k}).
\label{eq: energy_time}
\end{equation}
Because the system is also degenerate with respect to spin [since the operator ${\bf S}_i$ commutes with ${\mathcal H}(t)$], we find $\epsilon_\uparrow({\bf k})=\epsilon_\downarrow({\bf k})$, which then implies Eq.~(\ref{eq: energy_parity}). Finally, we also examine what happens in the presence of spin-orbit coupling
of atomic origin.  In this case, since the spin-orbit potential has the
periodicity of the lattice, the eigenstates remain in the Bloch form,
but since each component of spin no longer commutes with the Hamiltonian,
the spin becomes entangled with the wavevector.  We continue to have a 
Kramers doublet, where the spin label in  Eq.~(\ref{eq: energy_time})
is replaced by a pseudospin. 
Because ${\bf k}\rightarrow -{\bf k}$ and $\uparrow\rightarrow\downarrow$ 
under a time-reversal operation, the two degenerate states will have opposite velocities. Hence, in equilibrium (${\bf A}=0$), the total current vanishes because the contributions from the states with opposite momentum and opposite
pseudospin cancel in the summation in Eq.~(\ref{eq: current}).

Now consider the case where Eq.~(\ref{eq: energy_parity}) holds (inversion symmetry or time-reversal invariance and no spin terms in the Hamiltonian).  Then, we can replace the exponential in Eq.~(\ref{eq: band_structure}) by a cosine, so we have
\begin{equation}
\epsilon({\bf k})=-\sum_{\boldsymbol \delta}t_{\boldsymbol\delta}\cos({\bf k}\cdot{ \boldsymbol\delta}).
\end{equation}
The field-dependent velocity then becomes
\begin{equation}
{\bf v}_{\bf k}(t)=-\frac{1}{\hbar}\sum_{\boldsymbol\delta}{\boldsymbol \delta}t_{\boldsymbol\delta}\sin\left ( \left [{\bf k}-\frac{e{\bf A}(t)}{\hbar c}\right ] \cdot{\boldsymbol\delta}\right ).
\end{equation}
Using the trigonometric summation formula, we can expand the sine function into the difference of products of sines and cosines with arguments of ${\bf k}\cdot{\boldsymbol\delta}$ and ${\bf A}\cdot{\boldsymbol\delta}$. Substitute this into the equation for the current [Eq.~(\ref{eq: current})] and recall that the bandstructure is an even function of ${\bf k}$.  This implies that the only term that survives in the summation is the term proportional to $\cos{\bf k}\cdot{\boldsymbol\delta}$, since the sine term is an odd function of the wavevector. This then yields
\begin{equation}
{\bf J}(t)=\frac{e}{\hbar}\sum_{\bf k}\sum_{\boldsymbol\delta}{\boldsymbol\delta}t_{\boldsymbol\delta}\cos({\bf k}\cdot{\boldsymbol\delta})\sin\left (\frac{e{\bf A}(t)\cdot{\boldsymbol\delta}}{\hbar c}\right )f(\epsilon_{\bf k}-\mu).
\end{equation}
After performing the summation over momentum, the current will be expressed as a constant amplitude for each ${\boldsymbol\delta}$, ${\bf J}({\boldsymbol\delta})$, multiplied by the sine function which contains all of the time dependence, and then summed over the neighbors. We manipulate the sine function as follows:
\begin{eqnarray}
\sin\left (\frac{e{\bf A}(t)\cdot{\boldsymbol\delta}}{\hbar c}\right )&=&-\sin[{\bf E}_0\cdot{\boldsymbol\delta}\cos(\omega_0 t+\phi)g(t)]\\
&=&-{\rm Im}e^{i{\bf E}_0\cdot{\boldsymbol\delta}\cos(\omega_0 t+\phi)g(t)}\\
&=&-{\rm Im}\Biggr \{ J_0[{\bf E}_0\cdot{\boldsymbol\delta}g(t)]\\
&+&2\sum_{n=1}^{\infty}
i^nJ_n[{\bf E}_0\cdot{\boldsymbol\delta}g(t)]\cos[n(\omega_0t+\phi)]\Biggr \}\nonumber\\
&=&2\sum_{n=1}^{\infty}(-1)^nJ_{2n+1}[{\bf E}_0\cdot{\boldsymbol\delta}g(t)]\nonumber\\
&~&~~\times\cos[(2n+1)(\omega_0t+\phi)].
\end{eqnarray}
Here, we use the general formula $g(t)$ for the slowly varying envelope function, and the symbol $J_n$ denotes the $n$th Bessel function. The Fourier transform of this will correspond to sharp peaks around the odd harmonics $(2n+1)\omega_0$, with the width of the peak determined by how rapidly the envelope function $g(t)$ decays in time. The amplitude of the peak can decrease rather rapidly, as the Bessel functions decay for large index, with a fixed argument.  Hence, one expects to generically see the odd harmonics in the high harmonic generation 
if the system obeys the relationship in Eq.~(\ref{eq: energy_parity}). 
It is straightforward to generalize this argument to non-inversion-symmetric 
systems with spin-orbit coupling if the total current is simply the 
sum over contributions from the two bands with different pseudospin.

\begin{figure}[htb]
\includegraphics[scale=0.45]{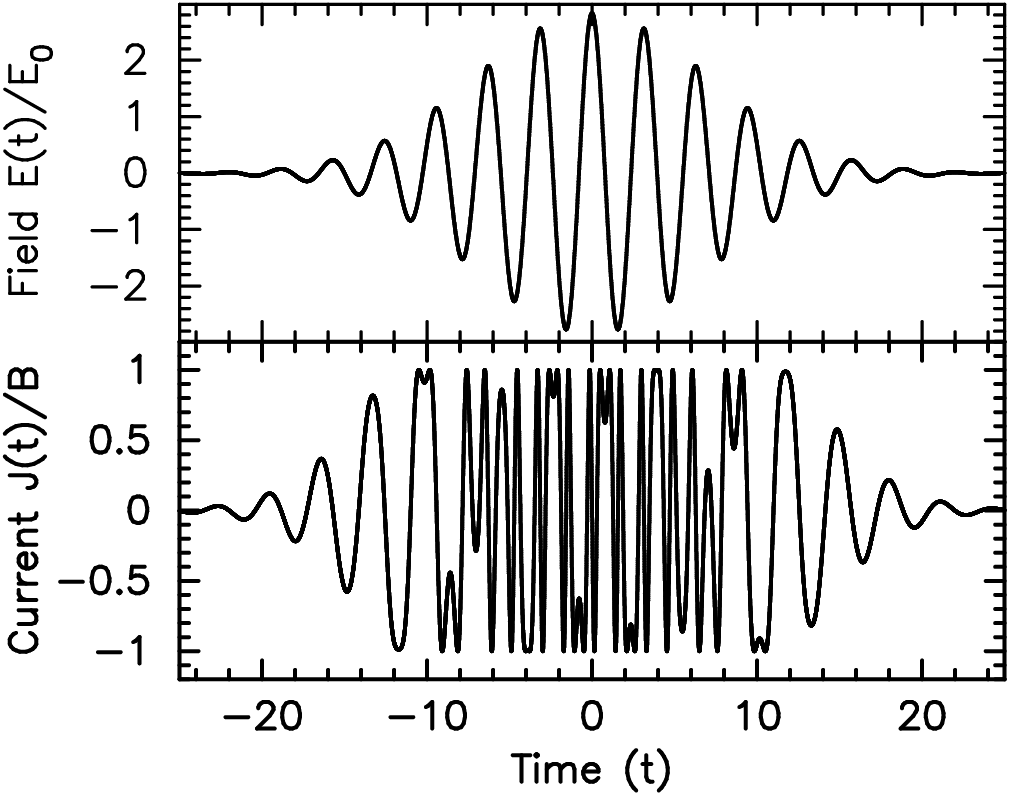}
\caption{Top panel: normalized time-dependent electric field with $\omega_0=2$, $\phi=\pi/2$, and $t_0=10$ in units of $\hbar\omega_0/ea$.
Bottom panel: current when $E_0=10$ in units of $B$.}
\label{fig: field_current}
\end{figure}

\section{Numerical results on model systems}

We will examine the generic current response for hypercubic lattices in $d$-dimensions, where the
field ${\bf E}_0$ has equal components along each coordinate direction, or has no component along a coordinate direction [hence we examine (1,0,0), (1,1,0), or (1,1,1) directions for a simple cubic lattice in $d=3$]. Using the fact that all coordinate directions are equivalent, so $\sum_{\bf k}\cos k_i a f(\epsilon_{\bf k}-\mu)$ is independent of the coordinate direction $i$ (with $a$ the lattice constant), then gives that the components of the current satisfy
\begin{equation}
J_i(t)=\frac{ea}{2\hbar d}\sum_{\boldsymbol\delta}\sin A_i(t)\int d\epsilon \epsilon\rho(\epsilon)f(\epsilon-\mu), 
\end{equation}
where $\rho(\epsilon)=\sum_{\bf k}\delta(\epsilon_{\bf k}-\epsilon)$ is the noninteracting density of states of the $d$-dimensional hypercubic lattice. This form of the current involves a temperature-dependent amplitude term $B=ea\int d\epsilon \epsilon\rho(\epsilon)f(\epsilon-\mu)/2d\hbar$ and the time-dependent term 
\begin{equation}
\frac{J_i(t)}{B}=\sin \left ( E_{0i}\frac{\hbar c}{ea}\cos(\omega_0t+\phi)\frac{1}{\sqrt{\pi}}e^{-(t/t_0)^2}\right ).
\label{eq: current_final}
\end{equation}
Here we define $E_0=A_0\hbar ct_0/ea$ which is the dimensionless amplitude of the electric field
(the units of the field are $\hbar\omega_0/ea$).
The radiated light is then proportional to $|\omega J_i(\omega)|^2$, which we find by taking the Fourier transform of Eq.~(\ref{eq: current_final}). Note that this specific form of the radiated harmonics is {\it independent of both the direction of the field and the dimensionality of the lattice for noninteracting electrons on hypercubic lattices}.

\begin{figure*}[t]
\includegraphics[scale=0.32]{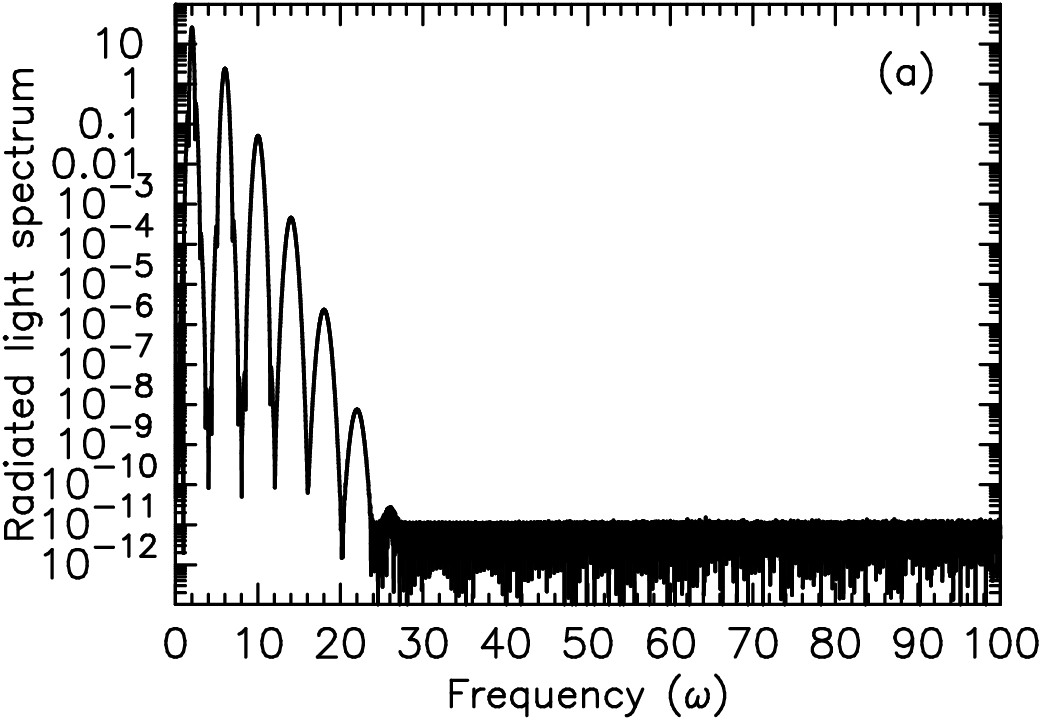}
\includegraphics[scale=0.32]{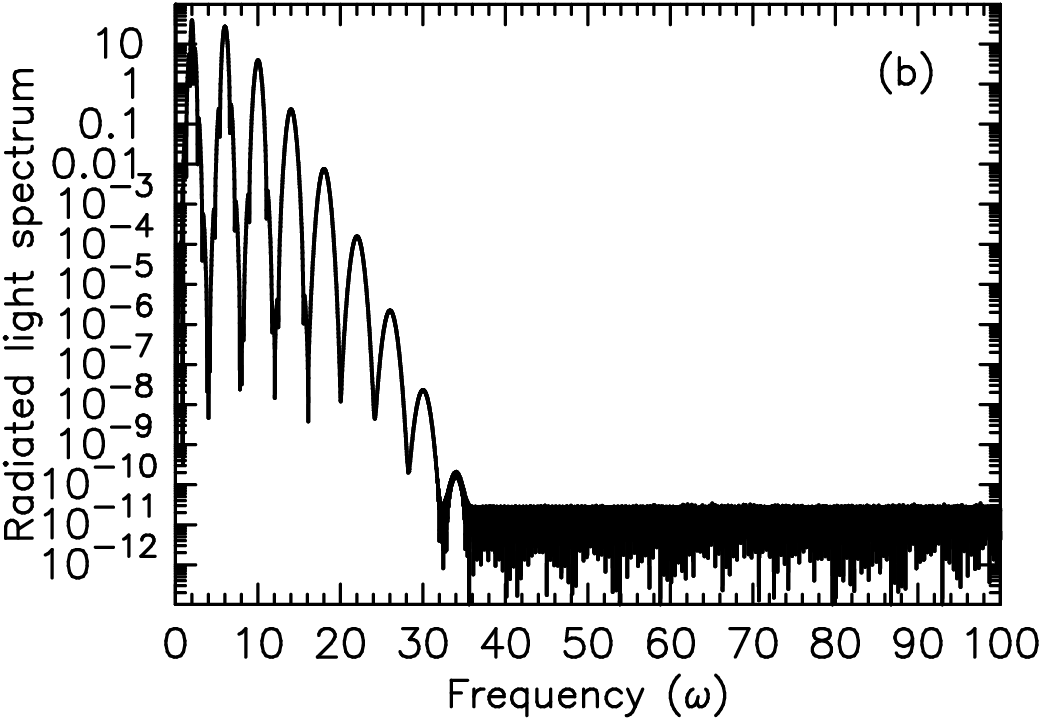}
\includegraphics[scale=0.32]{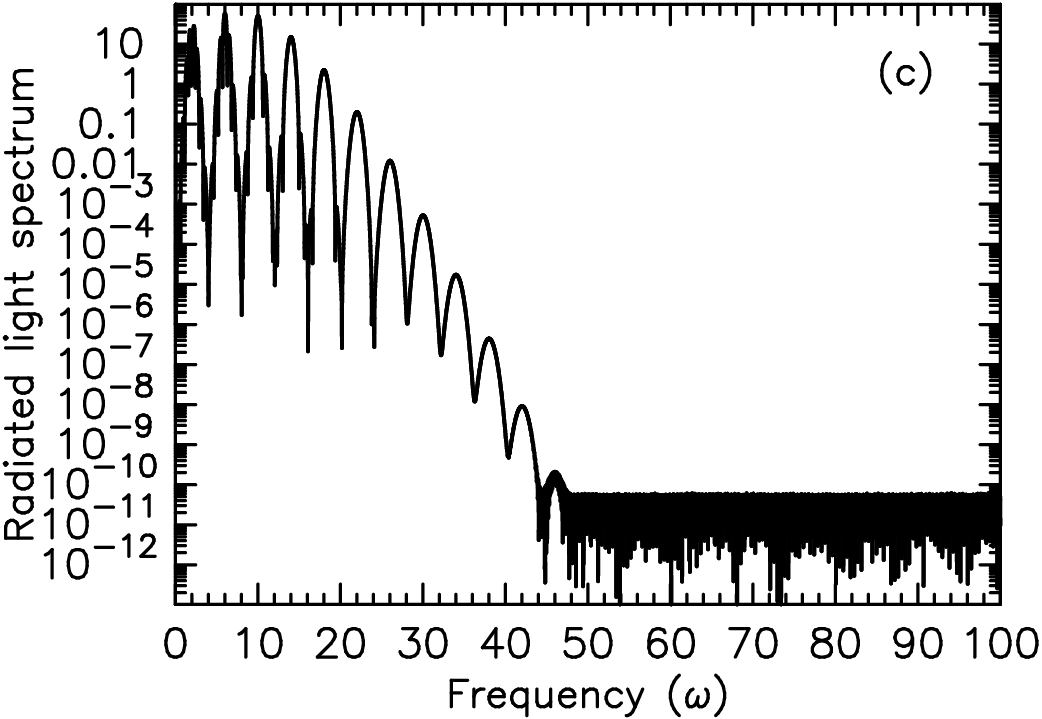}

\includegraphics[scale=0.32]{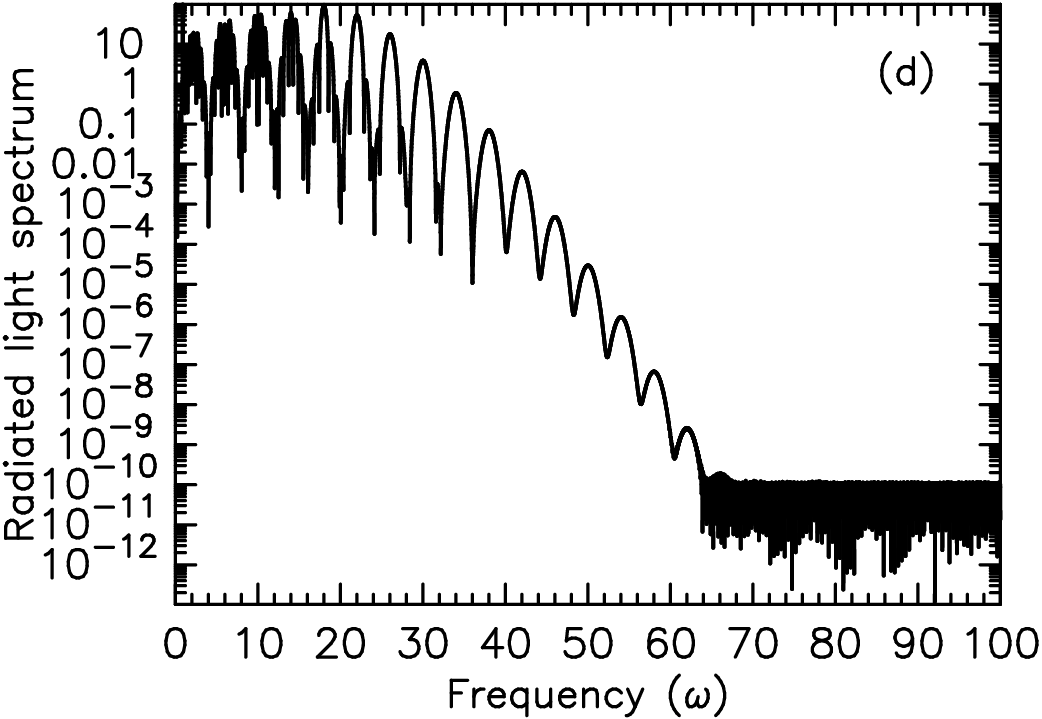}
\includegraphics[scale=0.32]{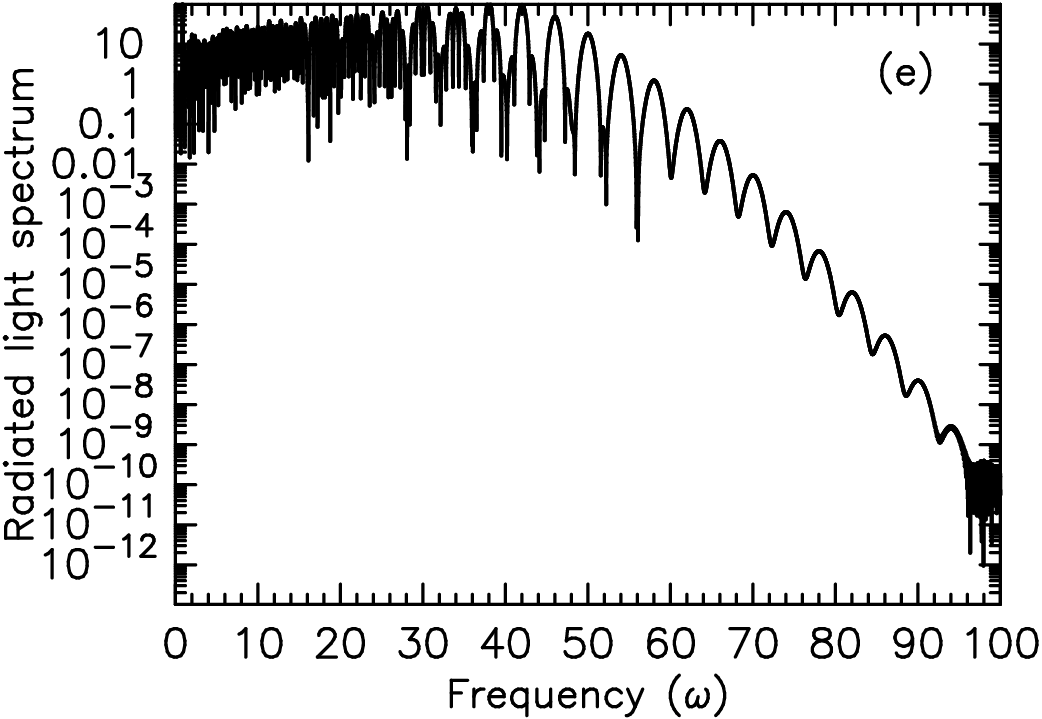}
\includegraphics[scale=0.32]{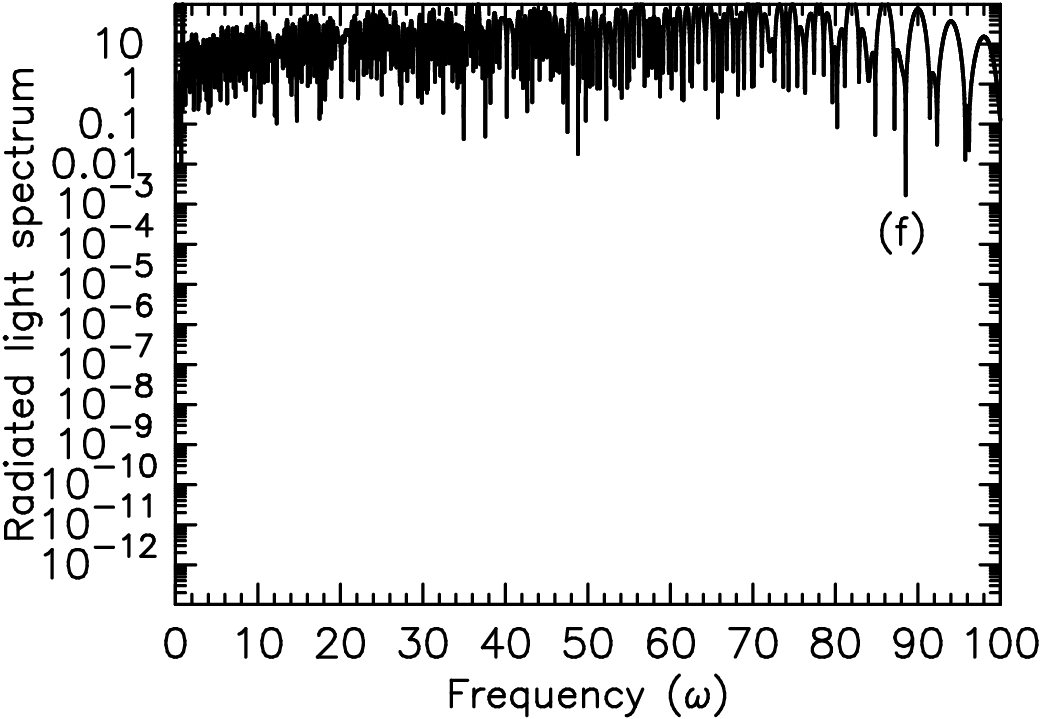}
\caption{Spectrum of radiated light from an electric field pulse with $\omega_0=2$, $\phi=\pi/2$, $t_0=10$, and $E_0$ taking on a range of different values: (a) $E_0=2$, (b) $E_0=5$, (c) $E_0=10$, (d) $E_0=20$, (e) $E_0=40$, and (f) $E_0=80$.  Note how the number of harmonics increases as the field amplitude increases, and how the signal appears to become noisier at small frequencies for higher field amplitudes.}
\label{fig: hhg}
\end{figure*}

We begin by showing the normalized electric field in the top panel of Fig.~\ref{fig: field_current}. We choose $\omega_0=2$, $\phi=\pi/2$, and $t_0=10$.  This produces an even (in time)  field profile that has on the order of 10-12 oscillations clearly visible, allowing for the frequency of the pump pulse to be reasonably well defined. The current as a function of time (without the temperature and dimension-dependent amplitude factor $B$) is shown in the bottom panel.  One can clearly see that as the field amplitude rises, the periods of the oscillations in the current get shorter, and that there is a ``clipping'' like effect in the oscillations when the field amplitude is large enough because the sine function is bounded between plus and minus one. The current has significant structure in it as a function of time and hence will give rise to a complicated Fourier transform.  But, by just looking at the results in Fig.~\ref{fig: field_current}, we can reason that we expect to see higher frequencies in the Fourier transform when the field amplitude is higher.  Indeed, this is what we next show.

In Fig.~\ref{fig: hhg}, we plot the logarithm of the radiated light spectrum in arbitrary units, as a function of the frequency of the light.  The field profile is identical to that found in Fig.~\ref{fig: field_current}, and the different panels show the effect of increasing the amplitude $E_0$ of the pump pulse. The plot is a semilogarithmic plot, because the higher harmonics are often suppressed by orders of magnitude from the fundamental before we hit a ``noise floor'' in the data.  Even though we have a completely well defined integrand to integrate over, and we use an adaptive integration routine available from quadpack for handling such Fourier integrals, it still is challenging to obtain high accuracy for these integrations.  The main source of error comes from balancing the integration range versus the order of the polynomials used in the subdivided integration ranges and the absolute accuracy requested in the subroutine.  We believe that most of our results with an amplitude larger than $10^{-10}$ are completely accurate, but we do not know whether there might be additional structure in the noise region.  Note that this is purely a numerical issue, and will have little bearing on experiment, because once the higher harmonics are too small in amplitude, they cannot be measured.  In general, we see on the order of a few up to maybe 20 higher harmonics appearing clearly in the figures.

\begin{figure}[bt]
\includegraphics[scale=0.45]{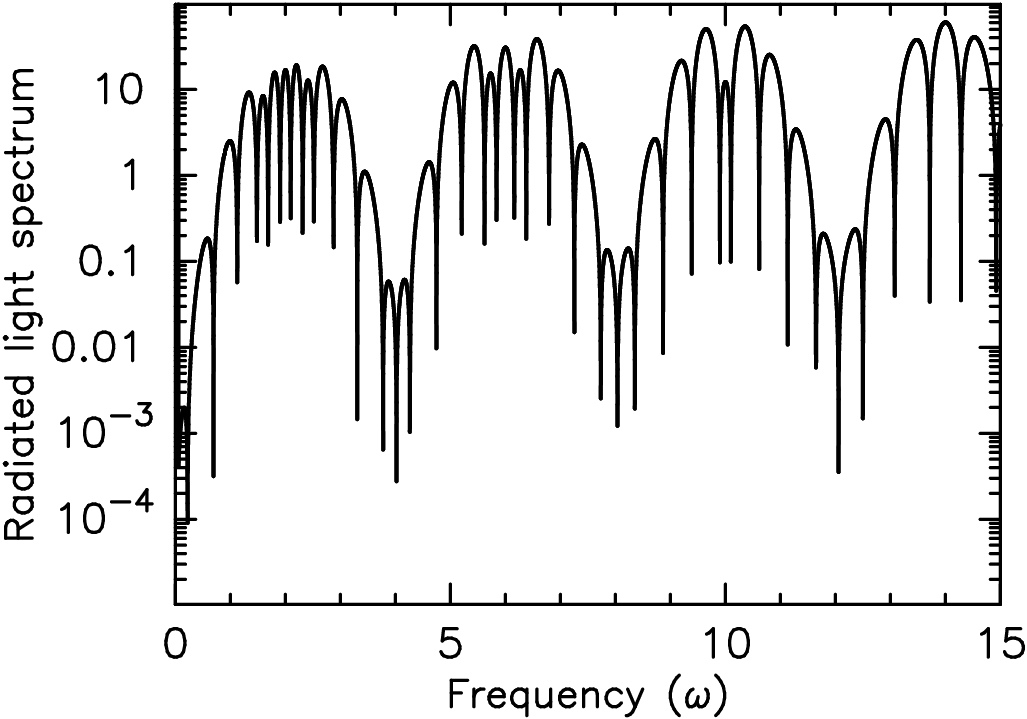}
\caption{Fine structure for low frequencies when $\omega_0=2$, $\phi=\pi/2$, $t_0=10$, and $E_0=20$ (magnification of panel \ref{fig: hhg}~(d) above). Note how the extra structure becomes less frequent for higher frequencies.}
\label{fig: hhg2}
\end{figure}

One can see two main trends occurring as the amplitude is increased: (i) first, the start of the noise floor moves to higher frequencies as the field amplitude is increased, as we expected, and hence we can see more higher harmonics as the system is driven harder, and (ii) we see first the development of a flat plateau of high amplitude harmonics occur at lower frequencies, which increases in size as the field amplitude increases.  But, we also see the development of additional structure that looks like noise for these high field-amplitude curves.  To investigate this further, we have enlarged the low-frequency region of panel {\ref{fig: hhg}~(d) in Fig.~\ref{fig: hhg2} for the case with $E_0=20$. One can see that this fine structure is not noise, as it appears in the figures with the wider frequency range, but is instead a fine structure of additional oscillations in the harmonic peaks that show more structure the lower the harmonic frequency is. The structure is reminiscent of a beating effect, but we have not been able to completely track down its origin.

In Fig.~\ref{fig: hhg3}, we investigate the effect of the phase of the field on the high harmonic generation.  The effect is rather mild, as can be seen in the figure, where the different colors correspond to different phases, but the evolution is definitely not monotonic in the phase.  Since detailed field profiles and phases of the electric field are difficult to determine for a given pump pulse, this weak phase dependence makes the experimental results much more robust, since the most important aspects of the field are the pulse width, oscillation frequency, and amplitude.

\begin{figure}[bht]
\includegraphics[scale=0.45]{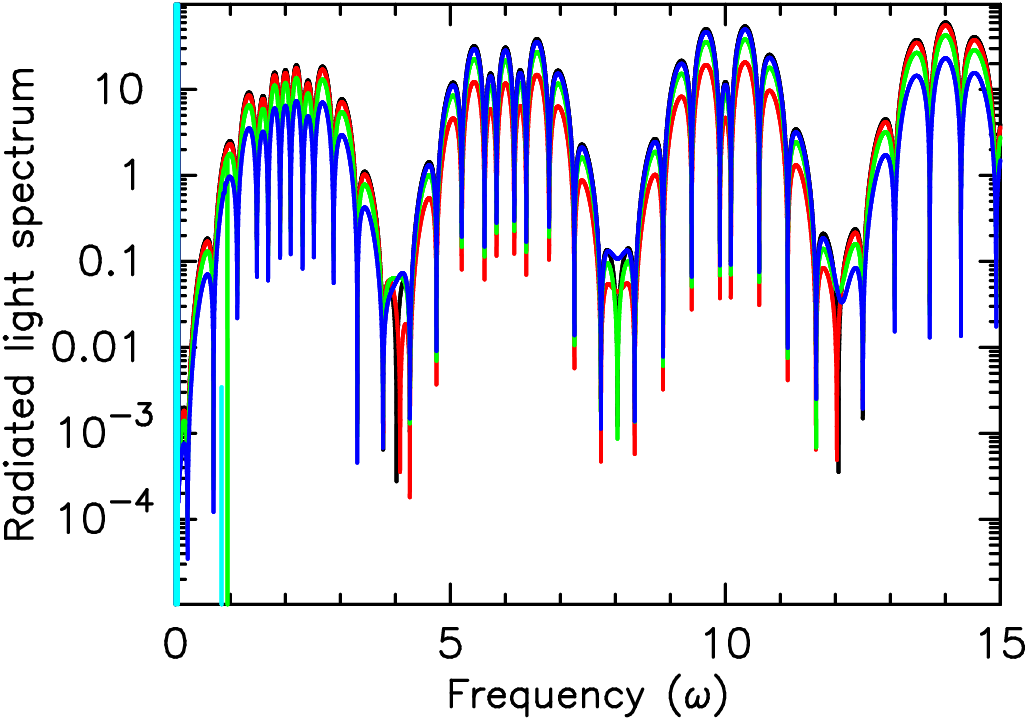}
\caption{(Color online) Effect of the phase $\phi$ on the fine structure. The different colors correspond to different phases: black ($\pi/2$); red ($3\pi/8$); green ($\pi/4$); blue ($\pi/8$); and light blue ($0$). Note how the dependence is nonmonotonic in the phase, but overall is quite mild. }
\label{fig: hhg3}
\end{figure}

\section{Numerical results for Zinc Oxide}

We now apply this model to
ZnO using the band structure calculated within density functional
theory (DFT). The wurtzite crystal structure of ZnO is uniaxial and
non-centrosymmetric, and spin-orbit coupling only weakly affects the
band structure (the splitting between bands at the top of the valence band
is only a few meV).  Thus
based on the symmetry arguments presented in Sec. II, we expect to see
only odd harmonics in the current response.

The DFT calculations were performed using the Vienna Ab Initio
Simulation Package (VASP) ~\cite{vasp}, with the electron-ion interaction
described using projector augmented wave potentials~\cite{PAW}, and the
exchange-correlation interaction handled with the generalized gradient
approximation (GGA)~\cite{PBE}.  Spin-orbit effects were not included. 
The structural parameters were optimized, and
the Kohn-Sham eigenvalues were tabulated on uniform grids in
the Brillouin zone. Band velocities were calculated using centered differences.
We have found that the noise floor in the calculated spectrum converges
slowly with respect to the Brillouin-zone sampling.  With a grid
of $64\times64\times32$ {\bf k}-points, 
the noise floor in $|\omega J(\omega)|^2$ 
lies about 7 orders of magnitude below the maximum,
which is sufficient to distinguish on the order of ten or more
harmonics, depending on the magnitude of the field.

The calculated band structure is shown in Fig. ~\ref{fig: BS}.
The highest lying valence bands have mostly O 2$p$ character,
while the conduction band is primarily of Zn 4$s$ character.
The GGA severely underestimates the ZnO band gap, yielding a
value of about 0.8 eV as compared to the measured gap of 3.4 eV.
Quasiparticle calculations based on the GW approximation
have found that the main correction to the DFT bands for ZnO is an upward
shift by about 2 eV of the two lowest conduction bands~\cite{vanSchilfgaarde}.
The overall shape and width of the O 2$p$ valence bands and of the
Zn 4$s$ conduction bands do not change significantly.  Since we address here 
only the contribution to high-order harmonic generation from field-induced 
acceleration of electrons within bands, and ignore contributions from 
field-induced dipole transitions between bands, the GGA bands provide a 
reasonable starting point for a band-by-band analysis. 

\begin{figure}[bt]
\includegraphics[scale=0.4]{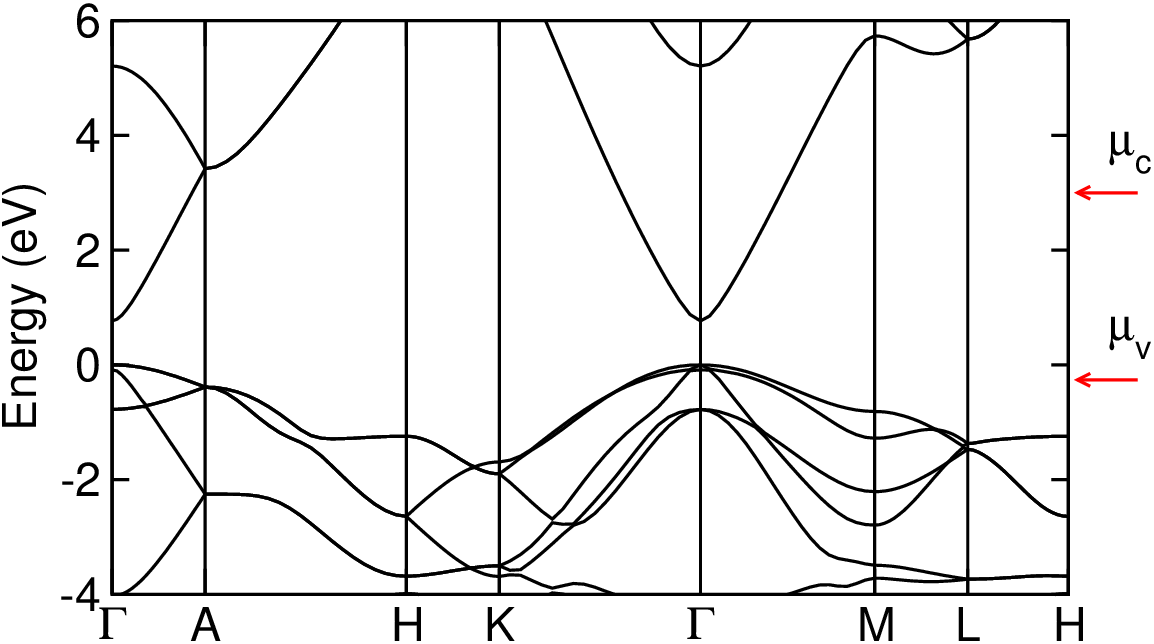}
\caption{ Band structure of wurzite ZnO calculated within GGA-DFT. 
The zero of energy is set to the top of the valence band. The
arrows indicate the chemical potentials
needed for a 10\% electron filling of the conduction band
and a 10\% hole filling of the valence bands.  }
\label{fig: BS}
\end{figure}

Because we are not modeling the photoexcitation process, 
we need to artificially shift the chemical potential to
create carriers in this semiconducting system. In the results presented
below, the response of electrons in the conduction band is calculated
by setting the chemical potential to the equilibrium value corresponding to a  
10\% occupation of the lowest conduction band. Given the large width of 
the Zn 4$s$ band, this corresponds to a chemical potential
about 2 eV above the conduction band minimum, as marked in
Fig. ~\ref{fig: BS}.  The O 2$p$ bands have larger
multiplicity and are much narrower than the conduction band, so 
a 10\% hole occupancy of valence bands requires a much smaller 
shift in the chemical potential with respect to the band edge. 
Without a mechanism for interband transitions, the response from 
holes is estimated by simply summing the current 
from all the valence bands with partial occupation.

A pulse of the form given by Eq. (8)  is applied, with 
phase $\phi=0$, width $t_0$ = 40 fs, frequency $\omega_0 = 20/t_0$, 
and maximum electric field strengths on the order of 0.1 to 1 V/\AA.
These parameters yield pump pulses similar to the 
$\sim9$ cycle, mid-infrared laser pulses used in  the recent 
experimental study of high-harmonic generation in ZnO \cite{zno_exp}.
Figure ~\ref{fig: hhgE0} shows the spectrum of light emitted 
by electrons in the conduction band when driven by an electric field polarized 
along the $[0001]$ direction.  The maximum field strength of
the pulse doubles between successive panels, showing the effect of
field amplitude on the spectrum.  The same general trends that were observed
in the calculation based on tight-binding bands are evident here. 
In particular, when the the field amplitude is increased, a plateau 
develops in the low-frequency regime, and the low-order
harmonic peaks acquire more and more fine structure. In the
experimental setup in Ref.~\cite{zno_exp} the detection
cut-off at low-energies was set approximately to the band gap, so the 
reported spectra start from about the 9th order harmonic and range up to
about the 25th harmonic.  Within this range,  the calculated decay of the 
intensity with increasing harmonic order agrees roughly with
the measured spectra. It would be interesting to try to extend the 
measurements to lower-order harmonics to see if the fine structure that appears
in our calculations is observed. 

\begin{figure}[tb]
\includegraphics[scale=0.4]{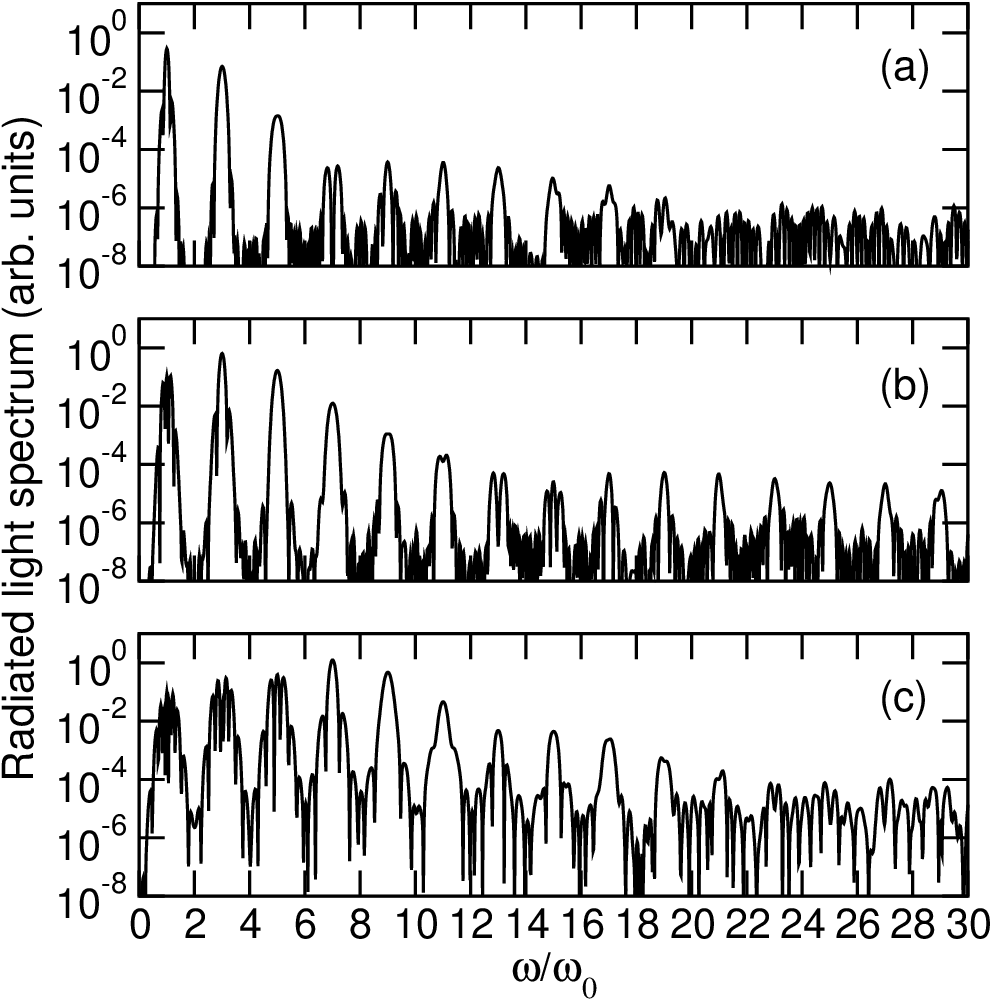}
\caption{Spectrum radiated by carriers in the ZnO conduction
band when driven by a pulse with the electric field polarized along [0001].
The maximum electric field strength is (a) 0.145 V/\AA, (b) 0.29 V/\AA, and 
(c) 0.58 V/\AA.  
}
\label{fig: hhgE0}
\end{figure}

Given the hexagonal symmetry of the wurtzite crystal structure,
it is expected that the response of electrons in ZnO will depend on 
whether the electric field is parallel or perpendicular to the $c$ axis.  
In addition, since the dispersion of the conduction and valence bands are
quite different, it is interesting to compare the emission spectra for 
electrons and holes.  Figure ~\ref{fig: hhgVB} shows the
spectra generated by electrons in the conduction band and
by holes in the valence band when driven by fields in the 
$[0001]$ and $[1\bar100]$ directions.  The parameters for the pump pulse 
(with the exception of the field direction) are the same as those used 
for Fig. ~\ref{fig: hhgE0}c.  When the driving field lies in the 
basal plane (Fig. ~\ref{fig: hhgVB}b and d),
the low-frequency plateau  region is shortened compared to when the field
is perpendicular to the plane (Fig.~\ref{fig: hhgVB}a and c).
The in-plane field produces a stronger signal for the lower harmonics, but
the magnitude of the harmonic peaks decays more quickly with harmonic
order. Further, the low-order peaks are
narrower and exhibit less structure than when the field is
polarized perpendicular to the plane. 
Despite significant differences in $\epsilon(\bf k)$ for the electrons
and holes, the spectral characteristics of their radiation are very similar.
There is an overall reduction in signal intensity from the valence band as
compared to the conduction band, which is consistent with the holes having
a signficantly larger band mass  than the electrons.  
Interestingly, the fine structure in the lower harmonics seems to 
depend much more on the direction of the electric field than  on
whether the carriers are in the valence or conduction band.  
At higher frequencies, the detailed shape of the peaks exhibits stronger
dependence on the band structure.  Varying the filling factors for the 
conduction and valence bands changes some of the detailed features in the 
spectra as well as the overall intensity of emitted light, but the general 
trends discussed here seem to be generic and robust.

\begin{figure}[tb]
\includegraphics[scale=0.33]{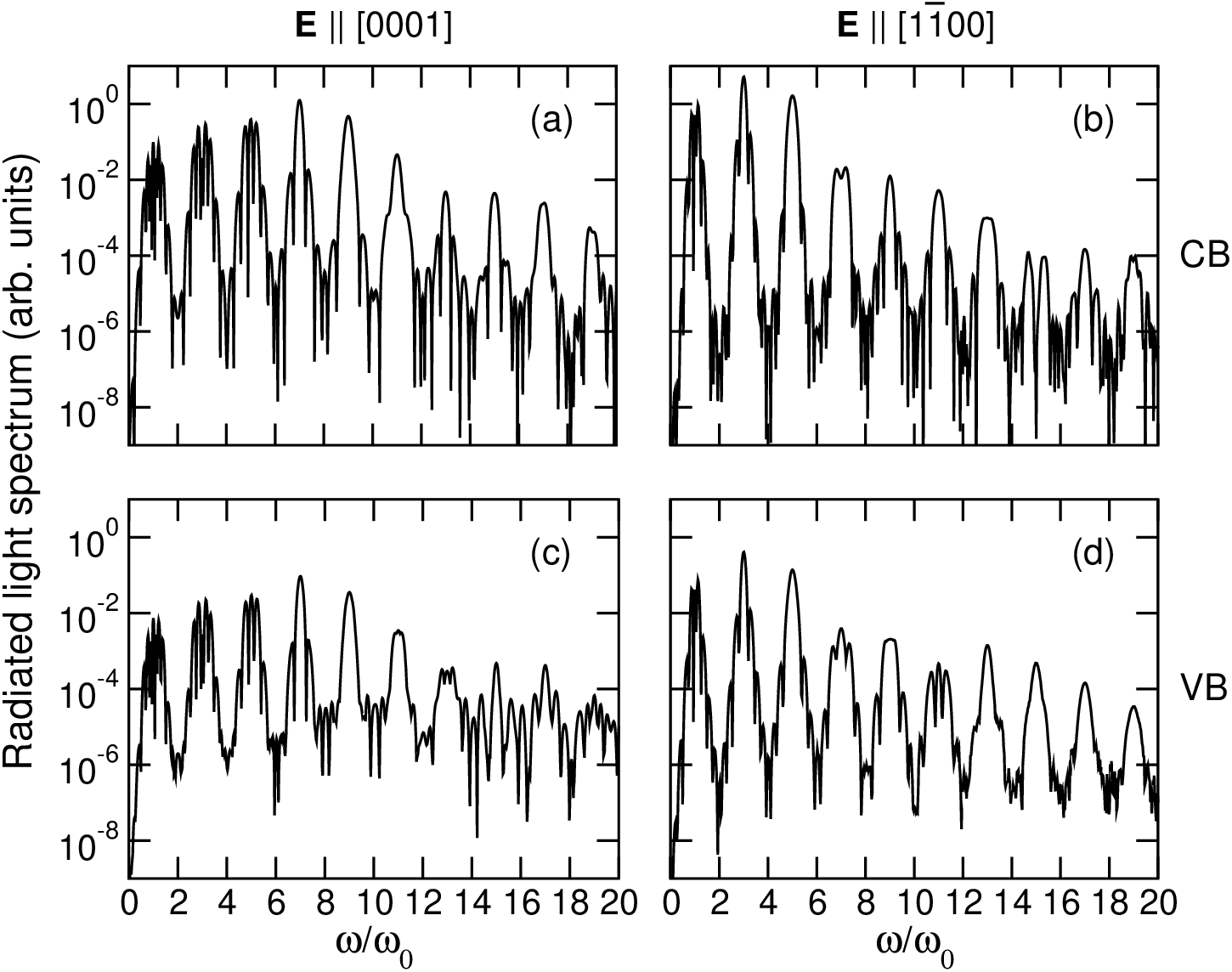}
\caption{Effects of field direction and band dispersion
on the emitted spectrum.  Panels (a) and (b) show emission
by electrons in the ZnO conduction band when driven by out-of-plane and 
in-plane electric fields, respectively.  Similarly, 
panels (c) and (d) show emission by holes in the ZnO valence band 
when driven by out-of-plane and in-plane fields, respectively. 
}
\label{fig: hhgVB}
\end{figure}

A remaining puzzle is that while most crystallographic
orientations produced only odd-order harmonics in the experiments, both
even and odd harmonics were observed when the polarization of 
the incident electric field was parallel (or close to parallel) to the
$c$ axis of the ZnO crystal \cite{zno_exp}.  Our model 
based on intraband dynamics of carriers in the bulk is not able to
account for the appearance of even harmonics under these circumstances. 
To more fully understand the experimental results for ZnO (or other
semiconductors), the coupling between bands~\cite{golde} and 
other effects such as  the broken symmetry at the surface need to be 
further explored. 

\section{Conclusions}
\label{sec:Conclusions}

In this work, we have shown some simple theories for high harmonic generation in metals (or photodoped semiconductors) which capture many of the experimental results already seen in this process.  While it has long been thought that Bloch oscillations occur at too fast a time scale to be observed experimentally in metals, this particular experiment is the exception, which allows one to indirectly observe the oscillations from the light generated by the accelerating electrons. Of course, in any real system, one needs to include interactions to determine how feasible the observation of these effects are.  The most important ones will be interactions with defects (disorder), with phonons, and with other electrons as the correlations in the material grow.  In order to do this, a more complete many-body physics formulation is needed on the Kadanoff-Baym-Keldysh contour~\cite{kadanoff_baym,keldysh}, which we have carried out an initial study for elsewhere~\cite{kemper}. It will be interesting to examine how this behavior survives as correlations increase, into materials like Mott insulators.  In addition, it will be important to understand the balance between the more universal behavior, which is captured by model system calculations, and the materials specific features, which require more materials specific calculations based on density functional theory plus extensions to include correlations in the nonequilibrium domain.  The future for these types of studies, both theoretical and experimental is clearly a bright one!
%%%%%%%%%%%%%%%%%%%%%%%%%%%%%%%%%%%%%%%%%%%%%%%%%%%%%%%%%%%%%%

\begin{acknowledgments}

The DFT calculations (AYL and JKF) were supported by the National Science Foundation under grant number DMR-1006605.
The model system calculations were supported by the U.S. Department of Energy, Basic Energy Sciences,
Materials Sciences and Engineering Division under contract No. DE-FG02-08ER46542 (JKF) and contract No. DE-AC02-76SF00515 (AFK and TPD).  JKF was also supported 
by the McDevitt bequest at Georgetown University. 
The Georgetown-Stanford collaboration was supported by
the U.S. Department of Energy, Basic Energy Sciences, Materials Sciences and Engineering
Division under contract No. DE-FG02-08ER46540.
\end{acknowledgments}

\end{document}